# Experimental and Theoretical Aspects of the Fragmentation of Carbon's Single and Multi-Walled Nanotubes

Sumera Javeed[1] and Shoaib Ahmad[2]*

[1]Pakistan Institute of Nuclear Science and Technology (PINSTECH), P. O. Nilore, Islamabad 45650, Pakistan. Email: subahyaqeen@gmail.com
[2]National Center for Physics, QAU Campus, Islamabad 44000, Pakistan
*Corresponding author: Email: sahmad.ncp@gmail.com

# Table of Contents

# List of Figures

## 1.1 Abstract

Irradiation with energetic ions is demonstrated as a powerful tool to investigate the fragmentation mechanisms of the single and multi-walled carbon nanotubes. The external irradiating ion's energy is consumed and dissipated in linear and nonlinear processes that are initiated in the nanotubes. The sputtered atoms and clusters that are emitted from the irradiated nanostructures are directly related with the energy dissipative and entropy generating dynamical processes for the generation of the defects. The multiple mechanisms of nanotube fragmentation are atomic collision cascades and localized thermal spikes that can be described separately with respective theoretical models. The experimental results from Cs-irradiated nanotubes are presented and explained using the atomic collision cascade and thermal spike models. An information-theoretic model is shown to comprehensively explain the diverse nanotube fragmentation mechanisms. The normalized experimental number densities of the sputtered species are shown to yield the probability distributions for every species emitted from the irradiated nanotubes. The model compiles Shannon entropy for each of the emitted constituent. Fractal dimension, defined and derived from Shannon entropy of the emergent species, is employed to determine the spatial dimension of atomic cascades and localized thermal spikes. Kullback-Leibler divergence or relative entropy utilizes the probabilities of emission of the sputtered species to identify and characterize the diversity of the mechanisms. The thermal and the information-theoretic models are used to identify, distinguish and characterize the existence and the relative operational efficiency of the collision cascades and thermal spikes in the irradiated, fragmenting single and multi-walled carbon nanotubes.

## 1.2 Introduction to the fragmentation of sp$^2$-bonded Carbon structures

Carbon nanotubes were discovered by Iijima while observing the transformation of the graphitic sheets under energetic electron irradiation in TEM [1]. The cylindrical hexagonal networks of C atoms curled in the form of nanometer-sized tubules were appropriately designated as carbon nanotubes CNTs. The single and multi-walled carbon nanotubes, SWCNTs and MWCNTs, were capped with the half spherical shells of fullerenes with the matching diameter. Fullerenes had already been independently discovered by Kroto et al few years earlier [2]. Pentagons were shown to induce spherical curvature in the spherical fullerenes. The newly discovered allotropes of carbon shared the ubiquitous sp$^2$ bonded C atoms. These C atoms are responsible for the formation of hexagons and pentagons during the synthesis of the fullerenes and carbon nanotubes [3]. The structural connectivity of carbon's allotropes was evidenced in the mechanisms of their formation and fragmentation. Thus the fragmentation of the sp$^2$-bonded graphitic sheets can be described as the essential, pre-formation stage. In Iijima's experiment, that was conducted with the energetic electrons having the energies E $\geq$ 100 eV in the transmission electron microscope, created the environment of the fragmentation followed by the topological rearrangements of the sheets into rolled nanotubes. CNTs were later formed by other techniques including the arc discharge between graphite electrodes [4,5]. In these techniques, the fragmentation mechanisms precede the formation of CNTs in carbonaceous discharges. Techniques for the formation of CNTs have been discussed in the preceding chapters of this Handbook and in various reviews and books, only a selected few of these are referenced as [6-9].

Prior to the discovery of CNTs, experimental results had shown that the mass spectrometric profiles of the subliming and evaporating graphite were dominated by the C clusters containing up to thirty (30) C atoms among the fragments [10-15]. A persistently observed feature emerging out of the mass spectra was the presence of the diatomic carbon molecule $C_2$. It was the most intense among the atomic and molecular fragments, especially the anions. Monatomic carbon $C_1$ was also present but had lesser intense peaks. The even number Carbon clusters were seen with comparatively higher intensities than the odd numbered clusters. These experimental observations and the respective interpretations on the structural form and nature of graphite's fragments existed prior to the discovery of CNT in 1991 [1]. When similar experiments were conducted with CNTs in place of graphite as the target, the experimental results of the sputtered clusters or the fragmented species showed similarities. Such a comparative mass spectrum is shown in Fig. 1.1 where SWCNTs of 2 nm diameter and 10-13 µm length compressed in 2 mm diameter x 5 mm thick Cu targets of the Source of Negative Ions by Cesium Sputtering (SNICS), were irradiated with $Cs^+$ ions at 5 keV [16].

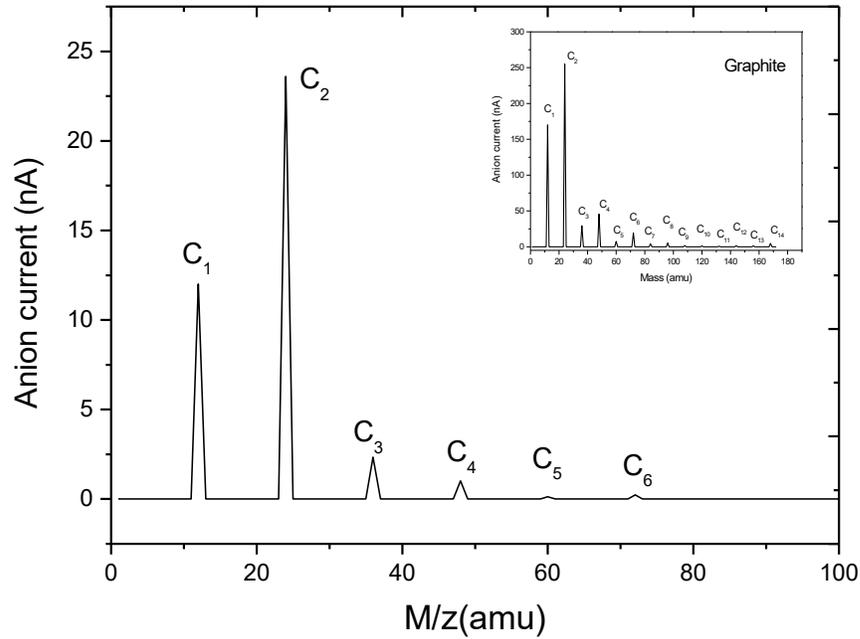

**Figure 0.1:** The mass spectra of negative sputtered atoms and clusters $C_x^-; x \geq 1$ from SWCNT irradiated with Cs$^+$ at 5 keV energy in SNICS. The **inset** has the sputtered species from graphite under identical conditions. The anionic currents are an order of magnitude higher for graphite as compared with those from SWCNTs. $C_2$ displays the highest intensity from both of the allotropes of carbon (from ref. [16]).

The discovery of CNTs and the continuation of the earlier sequence of experimental investigations of the fragmentation patterns and profiles have clarified various aspects of the underlying mechanisms. In this chapter, we will highlight the processes and mechanisms of the fragmentation of SWCNTs and MWCNTs. Experimental and theoretic aspects will be presented and discussed. The scheme of the chapter is the following: (1) Presentation of the results from the mass spectrometry of the fragments emitted from the irradiated SWCNTs and MWCNTs. (b) Discussion of the irradiation-induced damage's theoretical framework that separates the linear binary atomic collision cascades and the localized thermal spikes. (c) Presentation of the paradigm that the thermal, statistical mechanical and the information-theoretic models can be employed on the same data to illustrate the fundamental physical mechanisms. It will be demonstrated that the model developed by using the information-theoretic tools of Shannon entropy, the information-dependent fractal dimension and the relative entropy can provide a comprehensive description of the collision cascades (CC) and localized thermal spikes (LTS) as the emerging dynamical systems that are induced by the externally induced ions in CNTs.

## 1.3 The Fragmenting multi-walled CNTs:

MWCNTs have similarities with graphite for having the cylindrically bent graphene as coaxial cylindrical tubes with the inter-tube spacing almost similar to the inter-sheet spacing in graphite ~0.334 nm. The main difference appears in the dimensional characteristics of MWCNTs as compared to graphite where MWCNTs have one dimension (diameter) in the nanometer range while the lengths are in micrometers. Typically, a 10 nm diameter x 1 μm long MWCNT contains ~$10^8$ C atoms. The individual MWCNTs retain their structural identity even in a sample of the compressed MWCNTs and exhibit relatively poor conductivity as compared with the graphitic conductivity. The similarities and differences in their respective structural properties make MWCNTs and graphite as distinct and at the same time, comparable carbon structures for the study of irradiation induced structural defects in the form of sputtered species that are emitted from the targets.

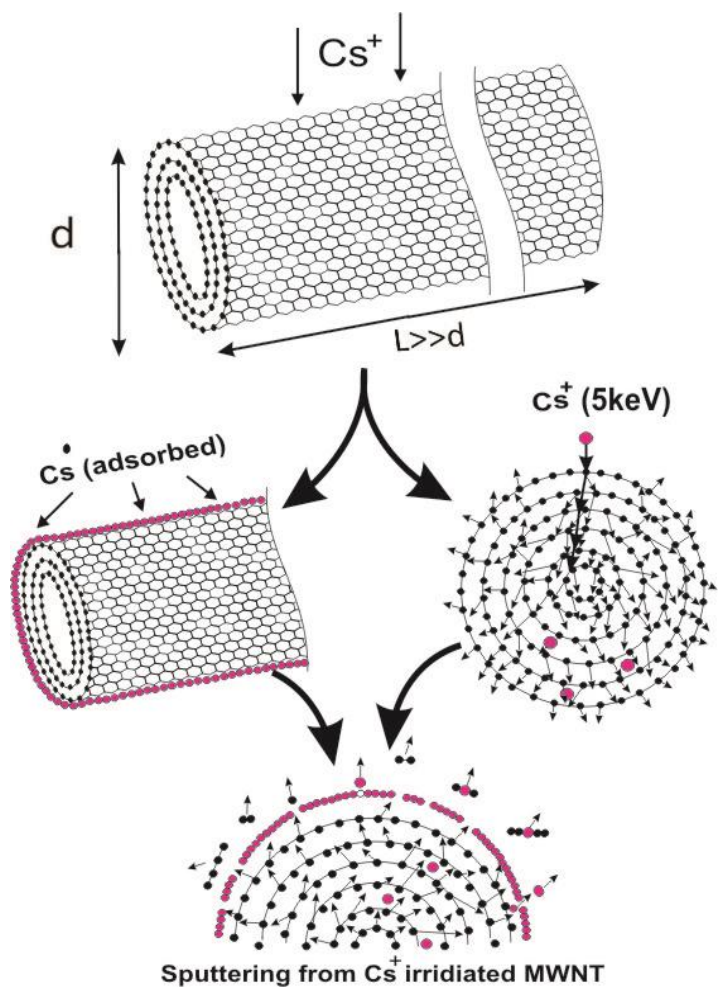

Figure 0.2: The schematics of $Cs^+$ irradiation of a MWCNT in SNICS.

It is assumed that the penetration ranges of the $Cs^+$ ions is less than the CNT diameter d while the length of nanotube L >> d as shown in Fig. 1.2. The typical irradiation-induced damage profile is shown in the figure with the initiation of collision cascades, creation of vacancies, generation of recoils and sputtering of the mobile atoms and clusters from the outer shell. Most sputtered species are neutral which acquire electron from the Cs-atomic layer on the nanotube. These are later extracted and mass analyzed as shown in Fig. 1.3 [16]. The spectra from MWCNTs shown in Fig. 1.3 contains $C_1^-$, $C_2^-$, $C_3^-$, $C_4^-$ and low intensities of higher clusters as anions. The dominant emitted species is $C_2$ at all energies.

$C_1$ acquires competitive intensities in the higher energy ranges $E(Cs^+) \geq 3.0$ keV. In addition, $O^-$ and $OH^-$ were sputtered from the outer cylindrical surfaces.

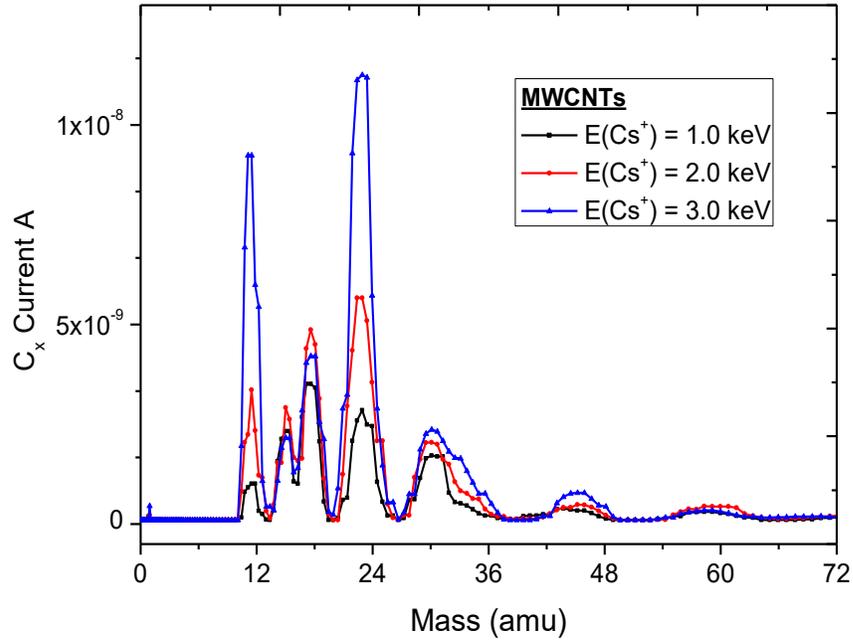

**Figure 0.3:** Three overlapping, mass spectra of the sputtered species $C_x^-$ at $E(Cs^+)$ =1.0, 2.0 and 3.0 keV.

In Fig. 1.4(a) the sputtering yields $Y_x$ of $C_1$-$C_4$ are plotted against $Cs^+$ energy $E(Cs^+)$. $C_2$ is the most intense sputtered species followed by $C_3$, $C_1$ and $C_4$ for the entire energy range. Fig. 1.4(b) plots the data of the normalized yields against $E(Cs^+)$ for all Carbon clusters. $C_2$'s relative yield decreases from ~ 100% to ~ 50% between low energy range 40 – 400 eV and then becomes stable for the higher energy range. The $C_3$ yield decreases from ~ 50% to ~ 20% up to 3.5 keV. The $C_1$ yield shows a gradual increase from ~ 5% to 20 % while $C_4$'s contribution is stable around ~10% for entire range of $E(Cs^+)$.

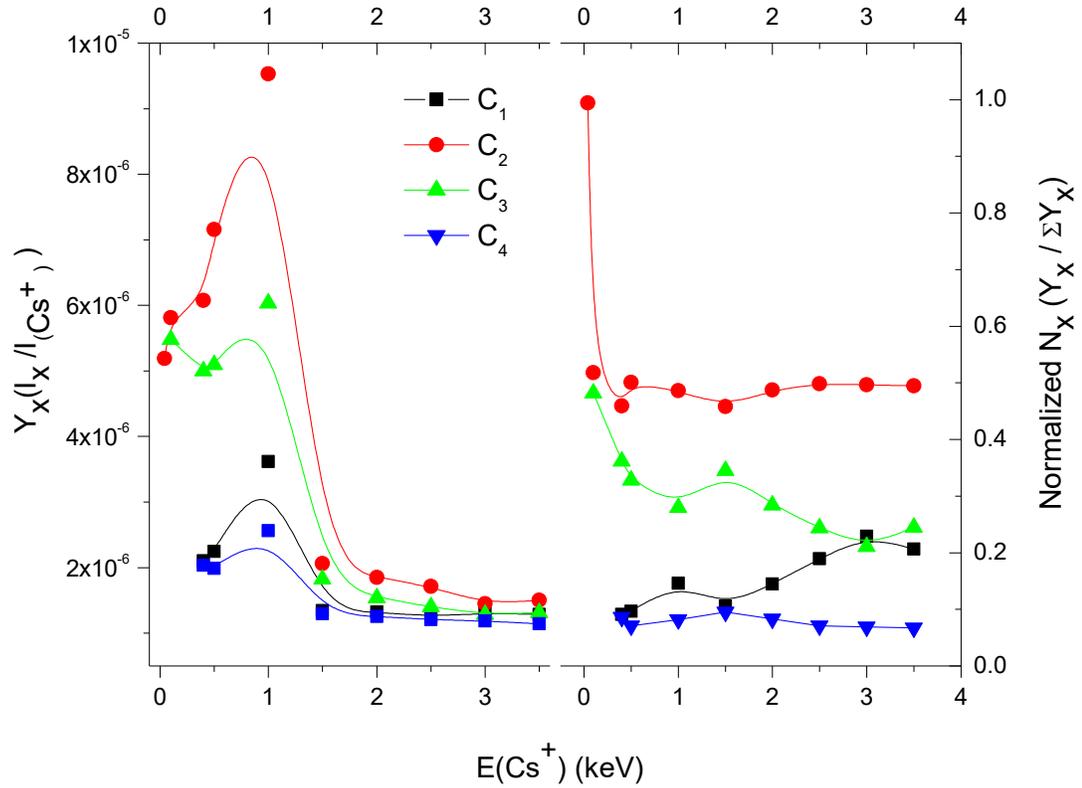

**Figure 0.4:** (a) The sputtering yield $Y_x$ and (b) normalized $N_x$ yields for $C_1$-$C_4$ are plotted for the pristine MWCNTs as a function of $E(Cs^+)$.

### 1.3.1 *Cumulative $Cs^+$ Induced Damage in MWCNTs:*

MWCNTs and graphite powder samples were irradiated for extended periods at high dose of $Cs^+$ at constant value of $E(Cs^+) = 5$ keV [16]. The same $E(Cs^+)$ range was selected to study the comparative, post irradiation effects in the two allotropes of carbon. The adsorption of neutral Cs on the surface and the implantation of the energetic $Cs^+$ play an important role in the formation and the emission of the carbon clusters ($C_x$; $x \geq 1$) and the cesium-substituted clusters (Cs-$C_x$) as anions. The associated structural changes are observed by using scanning electron microscopy.

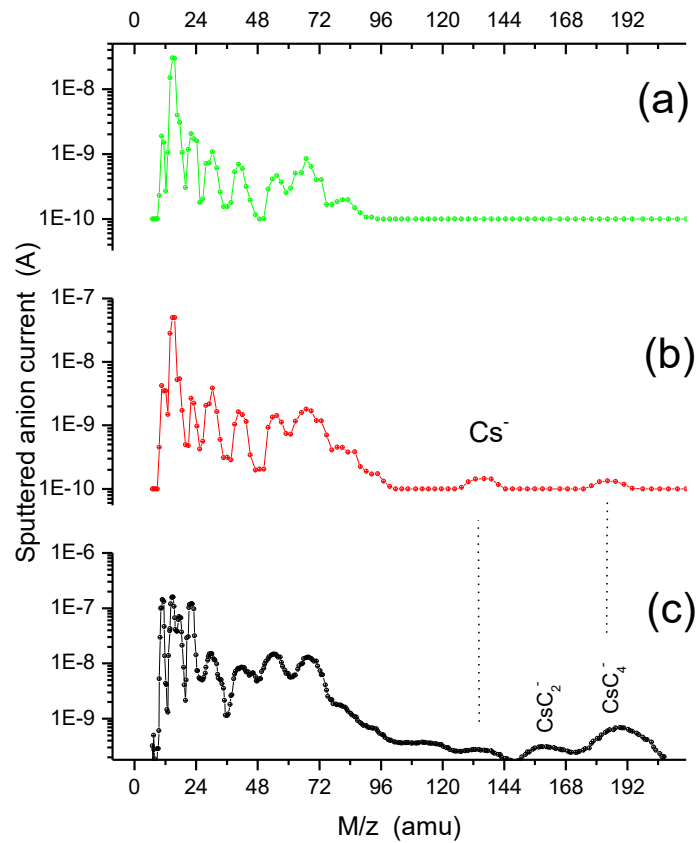

**Figure 0.5:** (a) The mass spectrum from pristine MWCNTs. (b) Mass spectrum after 20 min irradiation of MWCNTs. (c) The spectrum from a heavily irradiated MWNTs after 200 min of continuous $Cs^+$ irradiation.

Fig. 1.5 shows three mass spectra from the $Cs^+$-bombarded, compressed MWCNTs sample in the SNICS source [16]. The first spectrum 1.5(a) is from a pristine sample that was bombarded by $Cs^+$ beam for 5 minutes at 5.0 keV. The intensities are plotted on logarithmic scale to highlight the emergence of higher clusters ($C_x \geq C_2$) among the sputtered species as a function of ion dose. The spectrum shows monatomic, diatomic and higher C clusters up to $C_7$. At this energy the cluster intensities are low i.e. in the range of ~ nA. The sputtering of the adsorbed water-related peak of OH is also present with intensities higher than some of the $C_x$ anions. Here, it must be pointed out that the adsorbed water-related peaks of O and OH are a persistent feature of the sputtered species' spectra due to the omnipresence of water in nanotubes. Fig. 1.5(b) shows the mass spectrum collected after 20 min of irradiation at 5.0 keV. It can be noticed that the intensities of the carbon clusters $C_x$ are two to three orders of magnitude higher than those in the pristine sample. This spectrum has the first, clear evidence of the sputtered Cs⁻ that had been implanted during earlier rounds of irradiation. The cesium-mediated cluster $CsC_4$ is also visible. As pointed out earlier, water is absorbed during MWCNTs sample preparations and is very difficult to remove. However, it can be removed more effectively from graphite target than MWCNT depending on the nature and size of the two very different surfaces for the two carbon allotropes. Fig. 1.5(c) presents the spectra from a highly damaged sample of MWCNTs after 200 minutes of continuous $Cs^+$ bombardments with the cumulative dose of $10^{20}$ ions $m^{-2}$. Higher intensities ~10-100 nA of the C anions can be seen. $C_1$ to $C_6$ have well defined structures while the peaks of $C_7$ to $C_{10}$ are overlapped with the edge of $C_6$. Sputtered Cs can still be traced by the presence of Cs mediated clusters such as $CsC_2$ and $CsC_{4-6}$. Peaks from all three of the spectra are broad and overlapping. These are the hard-to-resolve high masses with our analyzer.

Irradiation of CNTs leads to the formation of both the pure carbon clusters $C_x$ and the cesium-substituted carbon clusters $CsC_x$ on the surfaces of MWCNTs and SWCNTs. At low doses, $C_1$s are more pronounced than $C_2$s from MWCNTs. Majority of sputtered species is neutral when leaving the nanotubes. The formation of anionic clusters occurs on the cesium covered surfaces by electron exchange between sputtered atoms and clusters with Cs atoms. MWCNTs provide larger areas as compared to SWCNTs due to the size variations. Experiments show that MWCNTs demonstrate pronounced structural transformation under comparative irradiation with $Cs^+$ ions. These may be due to the fact that those sputtered carbon clusters that may not escape outwards into the experimental chamber, may be deposited as surface adsorbents to act as fillers among the adjacent nanotubes in the case of MWCNTs. The re-deposition of the sputtered carbon clusters may participate in the localized melting and the formation of merged structures in MWCNTs. Localized melting due to thermal spikes will be discussed later in section 1.4. $Cs^+$ induced irradiation eventually may destroy the hexagonal structure of the surfaces of both types of CNTs leading to amorphous graphite. XRD spectra will be shown in the next section from the heavily irradiated SWCNTs and MWCNTs that provide the evidence of structural modifications.

## 1.4 Fragmentation profiles of the irradiated SWCNTs:

Experiments with the SWCNTs irradiated with $Cs^+$ ions as a function of energy and subjected to increasing $Cs^+$ dose, have confirmed similar results as were obtained from MWCNTs and identified a number of new aspects of the mechanisms of fragmentation of CNTs. It is because of the mono-shelled nature of SWCNTs that one can understand and visualize the operation of the mechanisms of CC and LTS by analyzing the mass spectra of the sputtered carbon atoms and clusters emanating from the irradiated surfaces. These spectra show the relative number densities of the fragmenting species and evolution of the accumulating damage due to the creation of vacancies. Irradiation induced fragmentation provides clues to the local structural transitions that occur. $C_2$, $C_3$ and $C_4$ are the most persistent sputtered species from the irradiated SWCNTs. $C_1$ has the least of the number densities among all species. It may be interpreted as the ease with which di-vacancies can be created as opposed to the single vacancies. The accumulation of the sputtered species in the inter-SWCNT space may be responsible for the increased connectivity among the nanotubes and result in the consequent increase in the electrical conductivity that was observed. The energetic ion irradiation effects have a destructive aspect that locally destroys the nanotube structure by introduces vacant spaces and at the same time the initiation of a reconstructive process that may occur due to the accumulation of the sputtered clusters.

Ion irradiations of SWCNTs, the consequent defect production mechanisms and the defect-related property changes have been investigated [17-19]. The relative ratios of the defects produced like the single versus di-vacancy and the role that displaced and knocked-off atoms, diatoms may play in the formation of new structures within and between the nanotubes has been simulated and experimentally studied [20-22]. Similar results from the irradiated $C_{60}$-fullerite has also identified the structural transitions that occur due to fragmentation of the $C_{60}$ cages as a function of $Cs^+$ energy $E(Cs^+)$ [23]. Prolonged $Cs^+$ irradiation led to the spherical fullerene cages' destruction. Similarly, sputtering profile of the clusters from the irradiated MWCNTs studies as a function of the ion dose had indicated the vacancies-related structural transformation [16]. Energy and dose of the irradiating ion, acts as a crucial factor that determines the nature and extent of the damage to carbon nanostructures. High doses of the heavy energetic ions may induce structural changes in nanotubes that vary from the fusion and welding of nanotubes to amorphous graphite [17,19,20,22]. Large scale sputtering from a collection of compressed carbon nanotubes may have twin effects; (1) single and di-vacancies in the nanotubes might open up the cylinders of tubes in localized regions and (2) the consequent re-deposition of these clusters on the neighbouring nanotubes could initiate new nanostructures in and around the nanotubes.

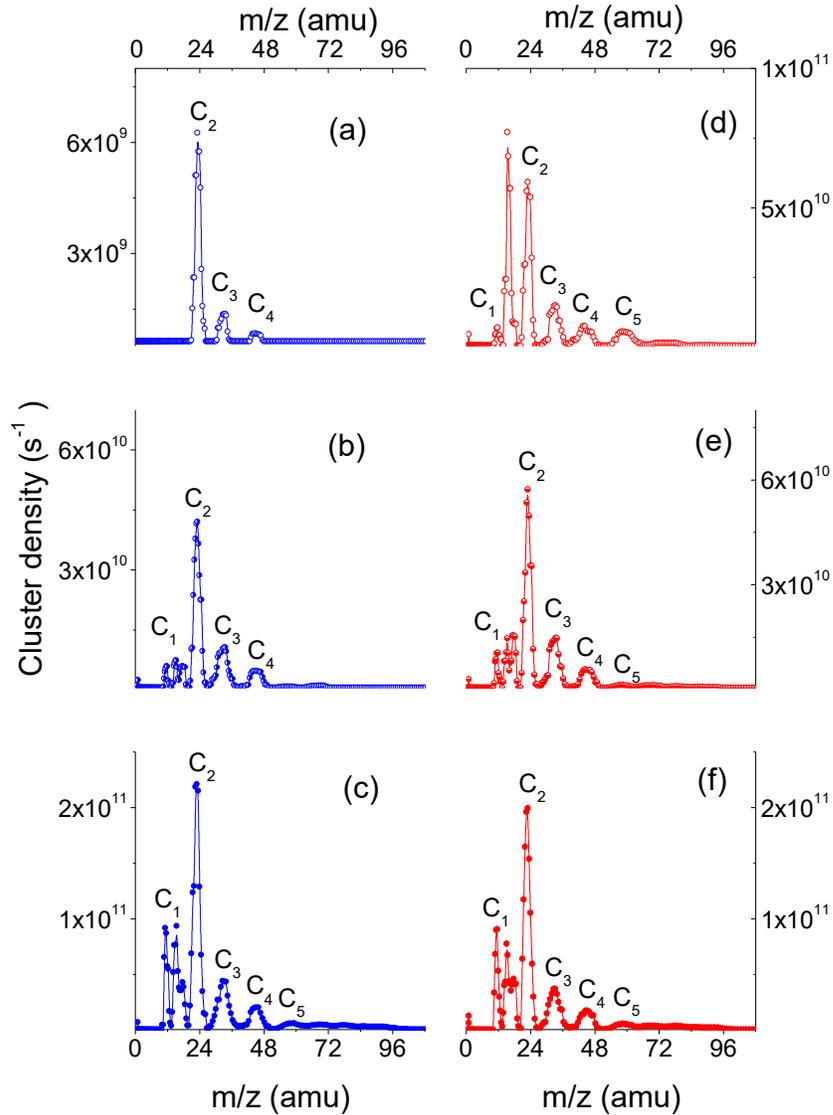

**Figure 0.6:** Two sets of mass spectra of $C_x$ sputtered from the irradiated SWCNTs are shown. The first set Fig. 6(a) to (c) shows $C_x$ emitted from the pristine sample while Fig. 16(d) to (f) are from the Cs rich, restructured SWCNTs at $E(Cs^+)$ = 0.2, 0.6 and 2.0 keV. Data from ref. [22,24].

Two sets of the mass/charge (m/z) spectra of the $C_x$ anions from SWCNTs were obtained as a function of $E(Cs^+)$. The spectra from the pristine sample are shown in Fig. 1.6 (a), (b)

& (c) at E($Cs^+$) energy 0.2, 0.8 and 2.0 keV, respectively. The other were repeat in the same energy range on the Cs-rich sample of SWCNTs. These results were chosen from the two sequences of nineteen m/z spectra of $C_x$ were obtained. Fig. 1.6(a) is the first mass spectrum from the pristine SWCNTs showing only $C_2$, $C_3$ and $C_4$. These three species retain the bulk of $C_x$ output for $Cs^+$ irradiations at all energies with relative variations in the intensities. $C_2$ is present in all the spectra as the most prominent fragment. $C_1$ makes its first appearance at E($Cs^+$) = 0.4 keV and steadily rise to about 10%.

### 1.4.1  *Normalized yields of the sputtered $C_x$ from irradiated SWCNTs:*

Fig. 1.7 shows the normalized yields designated as $C_x/\Sigma C_x$ for x=1 to 8 of the sputtered $C_x$ from the pristine SWCNTs. The normalized plots present the landscape of the comparative number densities. $C_2$, $C_3$ and $C_4$ are the main species emitted from the irradiated SWCNTs with the subtle differences between their number densities from the pristine set of SWCNTs in Fig. 1.7 and also those from the heavily irradiated ones in Fig. 1.8. These differences illustrate the nature of the damage to SWCNTs and the consequent emergence of the sputtered $C_x$-initiated structures. Fig. 1.7 shows the $C_2$'s relative yield decreasing from 64% at E($Cs^+$) =0.2 keV to ~50% between 1 and 2 keV. $C_3$'s yield increases from 20% to a broad peak of ~25% at 0.6 keV and stabilizes around its initial yield. $C_4$ appears as a relatively stable species with ~10% relative yield for the entire E($Cs^+$) range. The relative population of $C_1$ increases from 0 to ~10% indicating a uniform increase in the production of single vacancies with E($Cs^+$). The large clusters $C_5$ and $C_6$ are not detectable at E($Cs^+$) < 0.5 keV but once formed, these retain their steady share of ~3-5%.

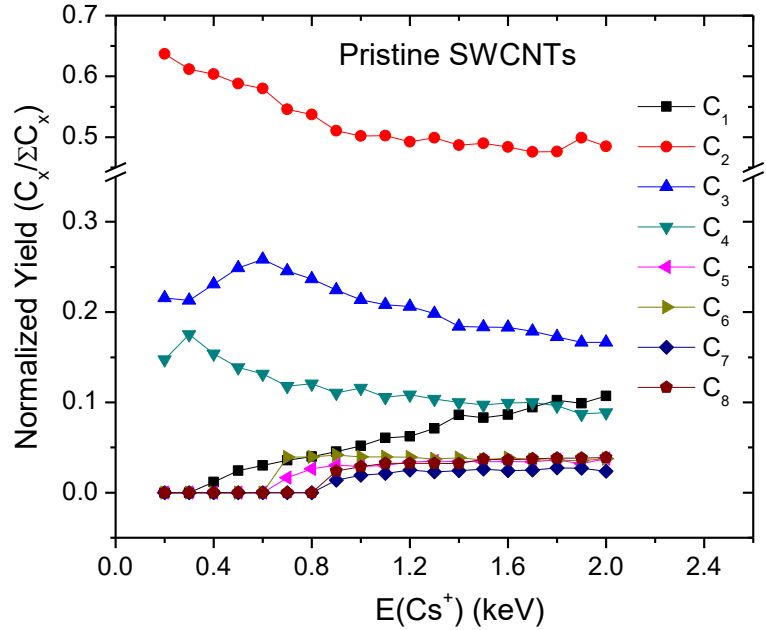

**Figure 0.7:** $\delta E(Cs^+) = 0.1\ keV$. The normalized yield $C_x/\Sigma C_x$ of $C_1$, $C_2$, $C_3$, $C_4$, $C_5$ …$C_8$ are plotted as a function of $E(Cs^+)$ from 0.2 to 2.0 keV for the pristine SWCNTs sample [24].

Fig. 1.8 shows similar broad feature of cluster emission profile that is obtained after 100 minutes' irradiation of the SWCNTs. The same ion energy range is chosen as in Figure 1.17. In the figure, the normalized yield graphs are shown for $C_2$, $C_3$, $C_4$ and higher clusters as the main constituents emitted from the $Cs^+$ sputtered nanotubes. The profile of $C_3$'s yield is similar to that in Fig. 1.7. $C_4$ also stays around a steady 10%. Monatomic $C_1$ can be seen not being emitted from the heavily irradiated ensemble of SWCNTs up to $E(Cs^+)$ =1.5 keV, after which its share is steady around 10%. This may indicate $C_1$ that as its yield is not proportional to the ion energy, as it was in Fig. 1.7, it could have been emitted as a by-product of the larger $C_x$ fragmentation.

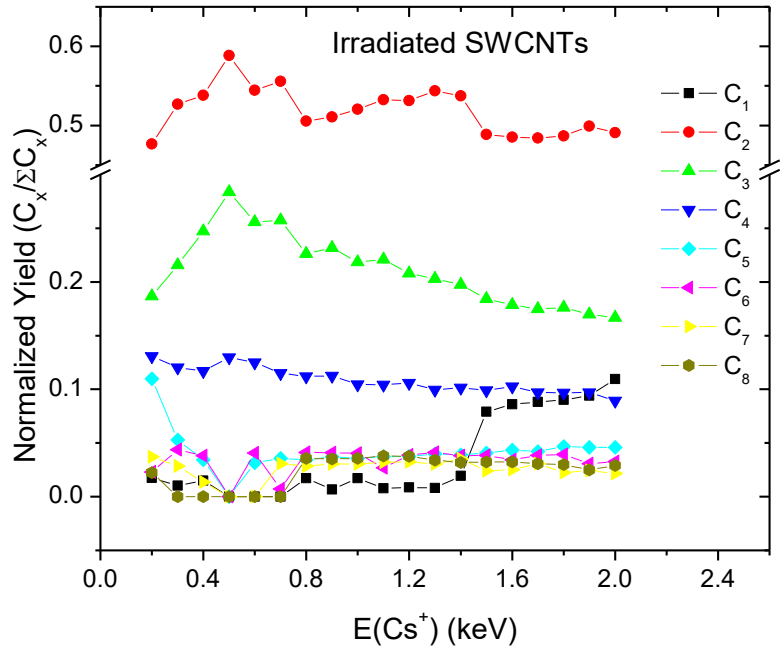

**Figure 0.8:** $\delta E(Cs^+) = 0.1\ keV$. The normalized yield $C_x/\Sigma C_x$ of $C_1$, $C_2$, $C_3$, $C_4$, $C_5$ …$C_8$ are plotted as a function of $E(Cs^+)$ from 0.2 to 2.0 keV for heavily damaged samples after Cs – irradiation [24].

Energy dissipating collisions of $Cs^+$ with SWCNTs has certain similarities with those in MWCNTs and also with those of the irradiated $C_{60}$-fullerite [23]. The mechanisms of the generation of primary recoils are similar in fullerite, and nanotubes, the relative yields of the collision products differs in each case. $Cs^+$-$C_{60}$ collisions in fullerite occur in different energy regimes; below 400 eV only $C_2$s are emitted while for the higher $E(Cs^+)$ all species are emitted [23]. This effect is unique to the irradiated fullerite and has not been observed in graphite, single or the multi walled carbon nanotubes. At the low irradiation regime, the irradiated SWCNTs throw out $C_2$ as the most intense species with $C_1$ being virtually absent; $C_3$ is seen as the next competing in intensity cluster with $C_4$ emitted having the third, stable cluster density. Collision cascades are likely to occur in MWCNTs with higher 3-D efficiencies of spreading as recoils among the overlapping nanotubes due to the availability of the nearest neighbors of a primary knock-on atom [25-28]. The spreading of the binary collision cascades does not seem likely to occur in SWCNTs, with the same efficiency as these may occur in MWCNTs. It is due to the lack of the adjacent shells which provide 3-D spreading of the cascades. The relative absence of $C_1$s in the mass spectra of irradiated SWCNTs and the associated single vacancies that ought to have been produced, provide the evidence. The observed predominance of the sputtered $C_2$s from all the C allotropes, demonstrates that the generation of di-vacancies are not unique to the irradiated SWCNTs and MWCNTs, these are also produced in $C_{60}$ and graphite, as discussed earlier. The diatomic $C_2$ emerges as the most significant fragmentation species from the irradiated CNTs. The physical mechanisms will be discussed and described in the next sections.

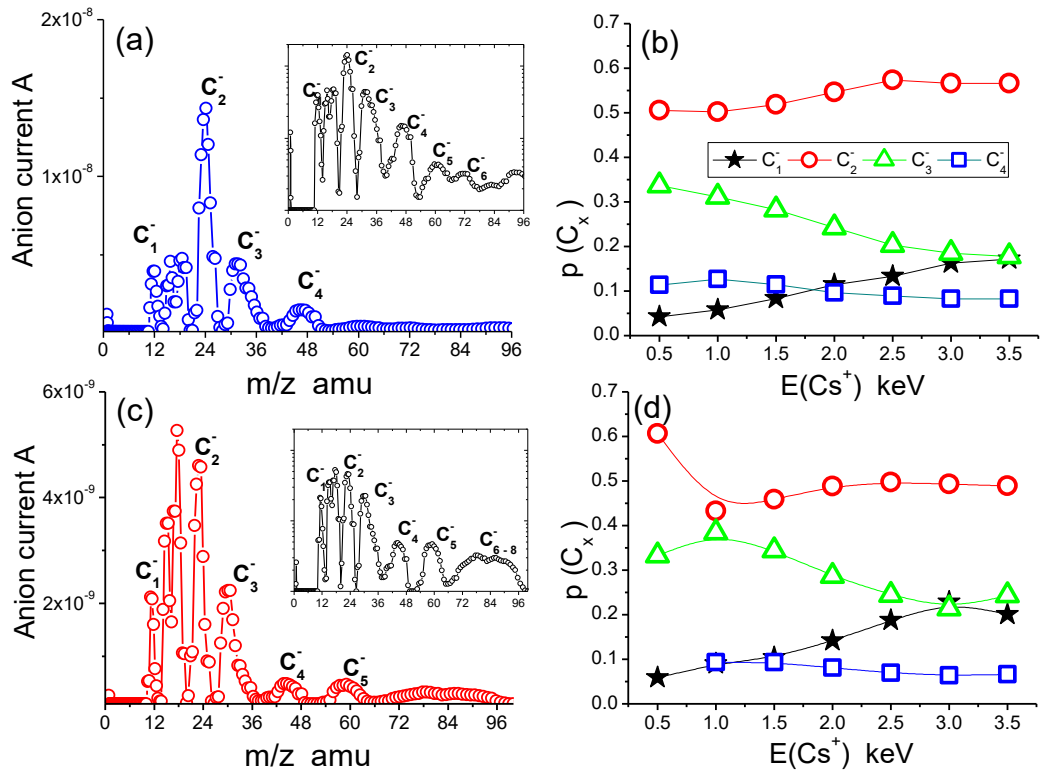

**Figure 0.9:** $\delta E(Cs^+) = 0.5\ keV$. (a) Mass spectrum of anions sputtered from $Cs^+$-irradiated SWCNTs at $E(Cs^+) = 1.5$ keV. **Inset:** shows the log-plot of the anion currents to demonstrate the presence and dominance of larger carbon clusters. (b) The probability of sputtering $p(C_x)$ of the four anionic species $C_1^-$, $C_2^-$, $C_3^-$ and $C_4^-$ are plotted as a function of $E(Cs^+)$. (c) Mass spectrum of sputtered anions from MWCNTs at energy

$E(Cs^+)$ = 1.5 keV. **Inset:** shows the log-plot of anion peaks. (d) The probability $p(C_x)$ for the four C species is shown for the same energy range as that of (b). Data from ref. [22].

Fig. 1.9 shows the experimental evidence of the atomic and cluster emissions from the two types of nanotubes with a coarse-grained measure of $\delta E(Cs^+) = 0.5\ keV$ [22]. It may be pointed out that the earlier figures 1.17 and 1.18 were obtained with $\delta E(Cs^+) = 0.1\ keV$. The range of E(Cs$^+$) is chosen from 0.5 to 3.5 keV. With this comparatively coarse-grained measure of $\delta E(Cs^+)$, one can expand the E(Cs$^+$) range to higher irradiation energies without excessive Cs$^+$-induced damage and avoid the effects of the Cs-implantation in nanotubes [16,24]. Fig. 1.9(a) and (c) show two representative mass spectra of anions emitted from SWCNTs and MWCNTs at Cs$^+$ energy E(Cs$^+$) =1.5 keV. In the Fig. 1.9(b) and (d), the probabilities $p(C_x)$ of the four sputtered species C$_1$, C$_2$, C$_3$ and C$_4$ as anions, are obtained from the respective normalized emission densities, and plotted against E(Cs$^+$). The figure clearly demonstrates the predominant emission of C$_2$, C$_3$ and C$_4$ while there is relatively lower probability for the emission of C$_1$. The cluster emission does not explicitly depend on ion energy while for C$_1$ the probability $p(C_1) \propto E(Cs^+)$; as one would expect from linear atomic collision cascade theories [25-28]. The non-dependence of $p(C_x); x \geq 2$ on Cs$^+$ energy variations has been discussed elsewhere as space-filling, multifractal, thermal spikes [29-30]. The thermal model for the generation of LTS will be discussed in the next section.

The accumulation of the sputtered species in the inter-CNT space may be responsible for the increased connectivity among the nanotubes and result in the consequent increase in the electrical conductivity [24]. Fragmentation of the SWNTs leads to the destruction of the nanotube structure locally leading to the opening of the tubes. The nature and extent of the cumulative irradiation induced damage is to fragment the individual nanotube structure while simultaneously bonding or welding these with each other. The energetic ion irradiation effects are destructive for the nanotube structure and at the same time a reconstructive process occur due to the accumulation of the sputtered clusters. The irradiation induced structural transitions occur with these two simultaneously operative processes.

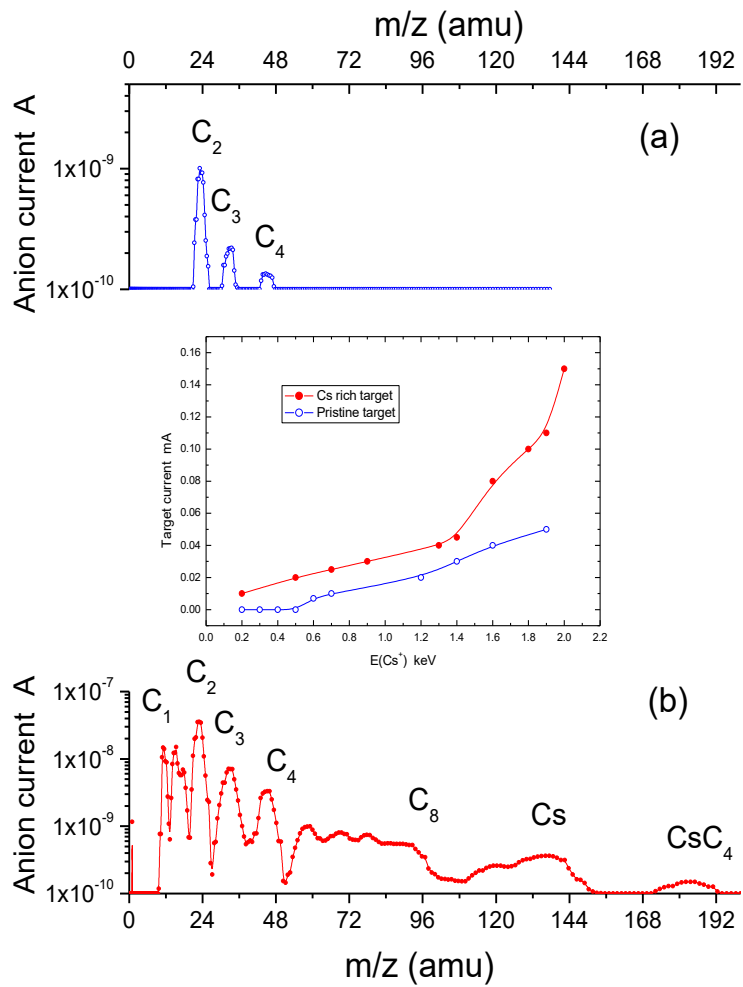

**Figure 0.10:** From pristine SWCNTs in (a) to the heavily irradiated SWCNTs (b). The implanted Cs peak is clearly marked along with CsC$_4$. (a) shows the first m/z spectrum with C$_x$ density on logarithmic scale to highlight the differences with those i.n (b) as the 38$^{th}$ spectrum at $E(Cs^+)$ = 2.0 keV, showing the entire array of C clusters C$_x$ (X=1 to 10), and broad peaks of Cs and CsC$_x$ (x=1, 4). **Inset** shows the increase in the target current as a function of the $E(Cs^+)$ and dose. The initial sequence on the pristine sample is shown with open and the one on the Cs-rich is shown with filled circles. Data from ref. [22,24].

The route of the aforementioned structural transformation that occurs while the nanotubes are losing their constituents, can be worked out from two experimental observations; one is the m/z plots for the first and the last spectra as Fig. 1.10(a) and (b), respectively. The inset shows the increase in the target current as a function of the $E(Cs^+)$ and the ion dose. The mass spectrum from the heavily irradiated SWCNT is plotted in Fig. 1.10(b). The anion current on logarithmic scale is used to enhance and clearly identify the low intensity large $C_x$ in the $38^{th}$, consecutive spectrum where fragmentation processes of the nanotubes are accompanied by the implantation of Cs into the CNTs. At low $E(Cs^+)$ and dose, only $C_2$, $C_3$ and $C_4$ are emitted, with $C_2$ being the dominant species. While, from a heavily irradiated sample of SWCNTs at high $E(Cs^+)$, all $C_x$ in the range $C_1$ to $C_{10}$ are sputtered, as shown in Fig. 1.10(b). $Cs^-$ as anion and Cs substituted carbon clusters $CsC_x$; x=1, 4 are also knocked off by the energetic $Cs^+$ ions.

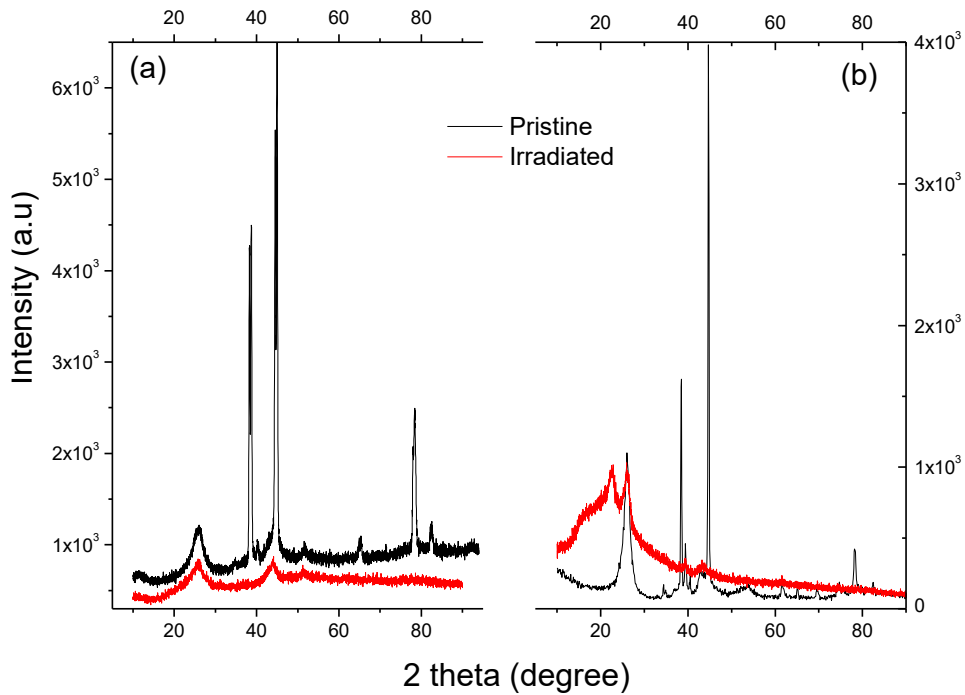

**Figure 0.11:** XRD spectra for (a) SWCNT and (b) MWCNT before and after irradiation with $Cs^+$.

Fig. 1.11 shows the XRD spectra from the pristine and the heavily irradiated samples of SWCNTs and MWCNTs. In both cases, the intensity of the peak corresponding to the main graphitic planes (002) at 26.2º decreased after irradiation and became broader, signifying the partial loss of the crystal structure. The substrate peaks are not shown in the irradiated samples because several layers of the irradiated carbon nanotubes were taken out from the holder and tapped on glass slide, XRD was then performed to avoid substrate peaks. In MWCNT spectra, the peak at 22.5º is from tape on the glass slide.

## 1.5 Thermal Origin of the emitted C Clusters:

Two fundamental questions arise when analyzing the mass spectra of carbon anions sputtered from $Cs^+$-irradiated single-walled carbon nanotubes:

1. Why is $C_2$ the main sputtered species rather than $C_1$?
2. Which mechanism explains the dominance of the multi-atomic carbon clusters' emission from the $Cs^+$-irradiated single-walled carbon nanotubes?

Collision cascade (CC)– based on binary atomic collision were introduced in the ion-induced sputtering of the solid targets' theoretical models [25-28]. These models, explained atomic sputtering but did not include or explain the sputtering of clusters. The irradiation-induced binary atomic CC in bulk solids evolve on and under the irradiated surface resulting in the sputtering of the atomic constituents. The primary events of the energetic projectile ion-target atom interactions initiate the cascades of recoiling atoms, generation of mono-valent vacancies where the energy is shared with the neighbours of the recoiling atoms. Sputtering yields are the ratio of atoms emitted per ion when the cascades intersect the surface. The sputtering models utilize the ion to target mass ratio, angle of ion incidence, energy of the incident ion as the input parameters. Ion energy is shared among the target recoils in binary collisions through appropriately chosen interatomic potentials [28]. Sputtering yield is the total flux of the recoiling target atoms that leave the outer surface per ion. Clusters are neither considered nor sputtered from the cascades. There is a direct dependence of the atomic sputtering yield to the deposited energy. However, the mono-shelled structure of the SWCNTs ensures that all outward expanding collision sequences are restricted to the same layer due to the monatomic layered nature of the nanotubes. The confinement of the spreading cascades to planar geometries implies the removal of the mobile atoms or clusters, as will be discussed next. Unlike the MWCNTs, where the mobile atoms or the clusters may reside in interstitial positions or within shells, SWCNTs ensure the removal of the sputtered atoms and clusters from the nanotube shell. Therefore, one observes rather well defined mass spectra of the sputtered species from SWCNTs. A thermal model was developed by considering and utilizing the earlier mentioned characteristics of the cluster emissions from irradiated SWCNTs [29] and was subsequently extended to the irradiated bulk solids like Si, Ge and ZnO [30].

### 1.5.1  *Localized Thermal Spike Model:*

A thermal model can be constructed when the emission of clusters could be assumed to be the net outcome of the emergence of a locally subliming region as a localized thermal spike (LTS). Such a model was built by utilizing the probabilities of emission of clusters that may be related with the formation energies of the respective multi-vacancies and the sublimation temperature $T_S$ [29,30]. The model considers a monolayer of N carbon atoms either of a planar graphene sheet or in the cylindrical form as a SWCNT. We assume that thermal spikes are initiated in localized regions that are at elevated temperatures $T_S$. It is further assumed that the monatomic or multi-atomic vacancies can be created if one, two, three, four or x- numbers of carbon atoms are removed in the form of clusters $C_x$. Single carbon atoms or clusters $C_x$ are bonded to the matrix of the surrounding C atoms with their respective binding energies $E_x$. To create $n$ vacancies with $x$ C atoms in a target with N carbon atoms, the number of ways this can be done is $W = \frac{N!}{(N-n)!n!}$. The associated entropy is $S_x = k \ln W$. The internal energy is $U = nE_x$ and temperature $T_s$ is related to $E_x$ and leading to entropy $S_x$ is $1/T_S = \frac{k}{E_x}(\partial W/\partial n)$. This leads to the ratio

$$p_x = \frac{n_x}{N_S} = \{(\exp(E_x/kT_S) + 1\}^{-1} = probability\ of\ emission\ of\ C_x \tag{1}$$

Here $p_x$ denotes the probability of creation of a vacancy of $x$-C atoms. The normalized yields of the sputtered clusters can be obtained from the consecutive experimental mass spectra of clusters $C_x$. The experimentally measured density for the emission of $C_x$ is directly proportional to the probability of the thermally created vacancies and is proportional to $p_x$. Probability of the emission of any of the cluster species, for example $C_2$, $C_3$, $C_4$ and the higher ones, is $p_x = n_x/N_S$, where $N_S$ is the total number of C atoms in the spike region. Since $N_S$ is an unknown quantity, we utilize the ratio of the probabilities of two clusters to eliminates $N_S$. Ratio of the probabilities allows calculation of physical quantities and variables like energies of formation of respective vacancies $E_{x,v}$ and the spike temperature $T_S$. These ratios are

$$p_x/p_y = \frac{n_x}{n_y} = \{exp(E_y/kT_S) + 1\}/\{exp(E_x/kT_S) + 1\} \tag{2}$$

In the first step of calculations, $C_1$ will be included in the application of thermal spike model along with the multi-atomic clusters. As its origin is not thermal and the density profile as a function of $E(Cs^+)$ is also different from those of the clusters ($C_2$, $C_3$ and $C_4$), therefore, such a calculation will only demonstrate the inapplicability of the thermal model to the emission of $C_1$. On the other hand, in Fig. 1.12, the ratios of the sputtering yields $C_2/C_3$ and $C_2/C_4$ are plotted against $E(Cs^+)$. These ratios clearly indicate the thermal nature of the emission profiles of the sputtered cluster species. The ratios of number densities of $C_2$-$C_3$ remain stable around 2.5±0.25 and 4.5±0.5 for $C_2/C_4$, for the entire range of Cs+ energies. These ratios of sputtering yields will be further used in calculating the spike temperature $T_S$.

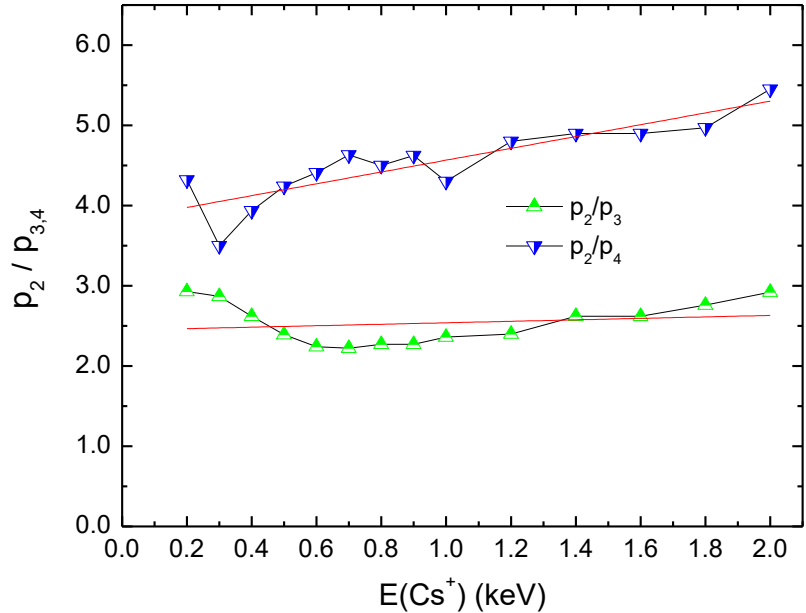

**Figure 0.12:** The ratios $n(C_2)/n(C_3)$ and $n(C_2)/n(C_4)$ are plotted as function of $E(Cs^+)$ [29].

## 1.5.2   *LTS Temperature $T_s$:*

The experimentally determined ratios for the probabilities of formation of vacancies $(p_x/p_y)$ along with the energies of formation of these vacancies in SWCNTs can be used to calculate $T_S$ by employing eq. (2). Energies of formation of single and di-vacancies are available in literature while the tri- and quarto-vacancies can later be evaluated by using the calculated values of $T_S$. The energies of formation of single and double vacancies $E_{SV}$ and $E_{DV}$ are obtained from DFT and DFT-based TB calculations for SWCNTs [31-34]. These are $E_{SV}$=6.8±1.0 eV and $E_{DV}$=4.7±0.5 eV. The first step is to use the computationally determined values of $E_{SV}$ and $E_{DV}$. Spike temperature $T_S$ is evaluated for each value of $E(Cs^+)$ by utilizing the proportionality $p_x \propto n_x$ in eq. (2) and the relation $p_x/p_y = n_x/n_y$. As $\exp(E_x/kT_S) \gg 1$, the expression for LTS temperature

$$T_S \cong [(E_{xv} - E_{yv})/k)][\ln(p_y/p_x)]^{-1} \qquad (3)$$

The calculated values of $T_S$ for the ratio $(p_2/p_1)$ for $C_2/C_1$ ratios are shown as open squares in Fig. 1.12 as a function of E(Cs$^+$). The calculated values of $T_S$ based on the ratio on $C_2/C_1$ vary between 3000 K at $E(Cs^+)$ =0.4 keV to 5500 K at $E(Cs^+)$ = 1.0 keV. Even higher values of $T_S$ ~ 9000 K are obtained for increasing $E(Cs^+)$. This line of argument demonstrates an energy dependent $T_S$ that produces increasingly larger values that are even higher than the boiling point of graphite. Such anomalously high temperatures are due to the use of the ratio of the probabilities of emission of two sputtered species $C_2$ and $C_1$ that have different origins and produced by two different physical mechanisms. $C_1$ in this case, is the outcome of the binary atomic collision cascades-a non-thermal process. $C_2$ on the other hand, is the outcome of thermal spikes in a localized region.

The average values of tri- and quarto-vacancies obtained at the temperatures obtained from $C_2/C_1$ are $E_{TV}$=5.13±0.5 eV and $E_{QV}$=5.73±0.5 eV for $E(Cs^+)$ = 0.4 to 1.0 keV. From the formation energies of $C_2$, $C_3$, $C_4$ and their experimental probabilities $p_2, p_3, p_4$, two sets of temperatures are calculated. These calculated values are shown in Fig. 1.13 as the filled red triangles and blue squares. A starred, black curve of the average of the two temperatures is shown $<T_S>$=4016K. This is the average spike temperature for $Cs^+-$ irradiated SWCNTs in the energy range from 0.2 to 2.0 keV.

The model developed using the experimental results of clusters emissions from $Cs^+$-irradiated SWCNTs can explain localized thermal spikes. Sputtering of clusters implied a thermal origin, and we have modeled it. The normalized number densities of clusters ($C_2$, $C_3$ and $C_4$) and their mutual ratios show constancy against the variations of the energy of $Cs^+$ as shown in Fig. 1.7. While $C_1$, in the same figure shows an energy $E(Cs^+)$ dependent behavior.

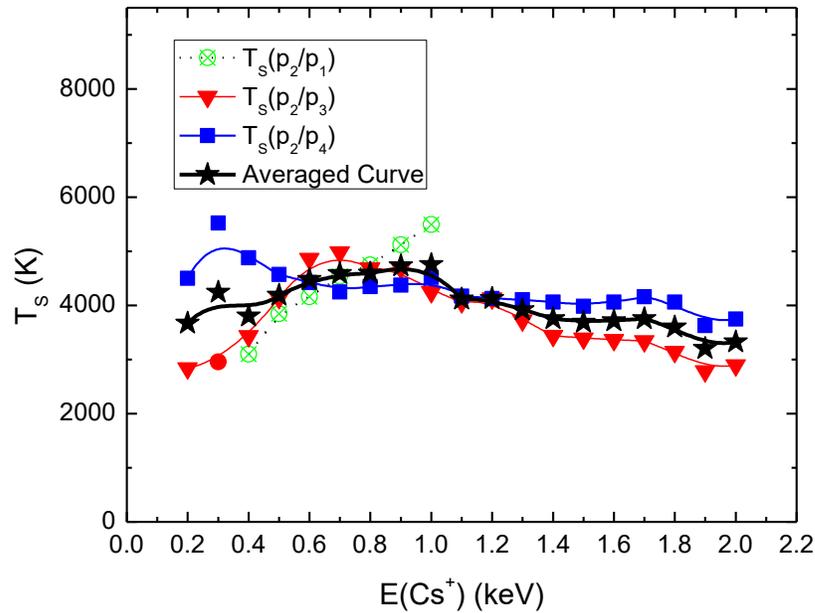

**Figure 0.13:** The spike temperature $T_S$ calculated from the normalized densities of $C_2/C_1$, $C_2/C_3$ and $C_2/C_4$ is plotted against $E(Cs^+)$. Average $<T_S>$ of $T_S$ ($C_2/C_3$) and $T_S$ ($C_2/C_4$) is shown as the thick black –starred line. Data from ref. [29].

## 1.6 CCs and LTSs as Information-Generating Dynamical Systems:

Solids irradiated by external energetic ions can also be treated as dynamical systems [35-37] where bifurcating recoil atomic trajectories may lead to collision cascades (CC) or localized thermal spikes (LTS). The two dynamical events are related yet occur in different time zones, as will be elucidated later. In this section, the dynamics of the emergence of these defects that have specific temporal and spatial profiles, will be used to characterize these as the entropy-generating dissipative structures. The dynamical system will be described by evaluating information or Shannon entropy [38] of all the dynamical processes. Experimental results from the irradiated carbon nanostructures [13-16,22-24] have demonstrated the need for revisiting the theoretical framework to explain the irradiation-induced damage on the 2-D surfaces of the SWCNTs. To identify the relative contributions of the mechanisms of CC and LTS, the normalized probabilities of emission of sputtered species are calculated from the experimental data of SWCNTs with 2 nm diameters, irradiated with Cs+ with $E(Cs^+)$ = 0.2 to 2.0 keV. From the experimentally determined probabilities in section 1.3.1 (Fig. 1.7), the instantaneous and cumulative entropies [38-40] of all sputtered species can be obtained. We show that the Renyi's definition of the fractal dimension [41] based on the cumulative information [42-44] generated, provides the dimensional evidence about the simultaneous existence and the relative contributions of the twin-mechanisms of the binary atomic collisions (CCs) and the thermal nature of the cumulative motion of multiple atoms as LTS. The fractal dimension of the four characteristic sputtered species ($C_1^-, C_2^-, C_3^-, C_4^-$) is used as a diagnostic tool that identifies CCs and LTSs and the dependent mechanisms of energy transfer by Cs⁺ ions to atoms of SWCNTs. It will be shown that the information-theoretic analysis is a powerful tool to probe the ion-solid interactions.

### 1.6.1 *The Probability Distribution Function:*

In this section, as in the earlier ones, the normalized probability mass distributions for each emitted species $p(C_x)$ has the same meaning and definition, and is evaluated at each increasing step of the irradiating Cs ion energy. Irradiating ion energy is treated as the variable parameter. The probability of generation of a lattice vacancy as a function of energy $E(Cs^+)$ had been defined earlier and once again for the formulation of Shannon entropy as

$$p(C_x) \equiv p_{C_x}(E(Cs^+)) \tag{4}$$

The advantages of employing this probability distribution function in the context of radiation damage theory, the information theoretic entropy and fractal dimensions is to develop a comprehensive model that helps us to understand and explain the four experimental observations from the mass spectra of sputtered atoms and clusters $C_{x\geq1}$. These are; (i) the relative absence of the monatomic $C_1$ at very low $E(Cs^+)$, (ii) the persistent cluster emissions with the higher, relative densities of $C_2$, $C_3$ and $C_4$ clusters at all irradiation energies, (iii) the energy dependent probability of emission of $p(C_1)$ and consequently of the single vacancy generation on the energy of the irradiating ions $E(Cs^+)$ and (iv) the observation that probabilities of the generation of multiple vacancies with the emission of $C_2$, $C_3$ and $C_4$ do not have explicit dependence on $E(Cs^+)$. These experimental observations were vividly demonstrated by the data in Fig. 1.6 (a) that has shown the mass spectra dominated by clusters $C_{x\geq1}$. Fig. 1.7 showed the cumulative data of emission probabilities $p(C_x)$ of the sputtered species of the entire energy range of $Cs^+$. The data presented in Fig. 1.7 is for the emission of anions from the irradiated, surfaces. The structural stability of the emitted anions is well established [45].

### 1.6.2 *Information Theoretic Entropy and Fractal Dimension:*

The information theoretic entropy $I_x$ can be evaluated from the experimentally determined probability distribution $p(C_x)$ of eq. (4) [38-40]

$$I_x = H(p(C_x)) = - \sum_\epsilon p(C_x) ln p(C_x) = \sum_\epsilon p(C_x) \ln\left(\frac{1}{p(C_x)}\right) \tag{5}$$

The summation is over all energy steps $\epsilon$ of the probability distribution for each sputtered species, from the minimum to the maximum. Two equivalent formulae are shown in eq. (5), the first with the negative sign to make sure that the negative values of $lnp(C_x)$ do not make entropy negative. The second formula with $\ln\left(\frac{1}{p(C_x)}\right)$ not only removes the need of the negative sign, it also introduces the increasingly important role of the function $\ln\left(\frac{1}{p(C_x)}\right)$ for low values of $p(C_x)$. The second notation for $I_x$ is the preferred one as it demonstrates the importance of the very low probability events. The information theoretic analysis can highlight all of the processes where the external energetic ions induce radiation damage in the irradiated nanotubes. Information or Shannon entropy $I_x$ for each sputtered species provides a measure of information about the underlying mechanism that is responsible for the emission of atoms and clusters. The output from the irradiated nanotube is treated as the information given out from the SWCNT that received a well-defined ionic energy input. The instantaneous values of information $p(C_x)\ln\left(\frac{1}{p(C_x)}\right)$ are plotted for the corresponding values of Cs energy in the figures below. The sum of the instantaneous information $p(C_x)\ln(\frac{1}{p(C_x)})$ yields $I_x$.

In Fig. 1.14(a) and (b), the radiation damage aspect of the model is demonstrated. The probabilities $p(C_x)$ for the pristine and damaged SWCNTs are derived from the data of anion currents obtained from the two sets of the nineteen (19) mass spectra each, with $\epsilon = 0.1\ keV$. Information-theoretic entropy $I_x$ is the sum of $p(C_x)\ln(1/(p(C_x))$ versus $E(Cs^+)$ shown in Fig. 1.14(c) and (d). These are obtained from the experimental normalized emission densities as the probability data for the pristine and the damaged sets of SWCNTs.

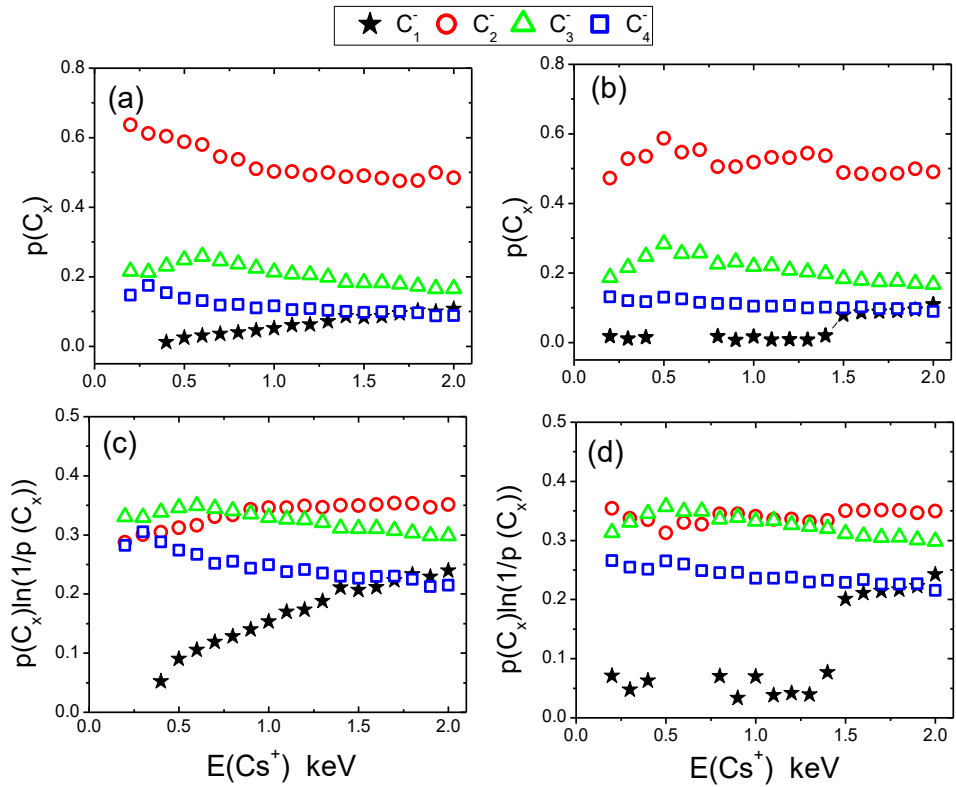

**Figure 0.14:** (a) The probability distributions $p(C_x)$, for the pristine sample of SWCNTs, of the 4 emitted species $C_1$, $C_2$, $C_3$ and $C_4$ are shown from the normalized anion currents at each step of $E(Cs^+)$. (b) For the highly irradiated and damaged set of SWCNTs, the probability distributions are plotted against $E(Cs^+)$. (c) The instantaneous information $p(C_x)\ln(1/(p(C_x))$ for each of the species, at every energy step $E(Cs^+)$ is plotted for the pristine sample with $p(C_x)$ obtained from (a). Figure (d) has the same for the highly irradiated SWCNTs as shown in (b). Data from ref. [40].

Information $I_x$ alone cannot provide conclusive evidence about the detailed character of the energy dissipation mechanisms. We show that fractal dimension derived from information $I_x$ emerges as an additional, analytical tool to unambiguously characterize the mechanisms responsible for atomic and cluster emissions. Fractal dimension can be calculated from the plots of $p(C_x)\ln(1/(p(C_x))$ against $E(Cs^+)$ as defined by Renyi [41]

$$d_f(C_x) = \sum_\epsilon p(C_x) \ln(1/p(C_x))/\ln(1/\epsilon) \qquad (6)$$

Here $\epsilon$ refers to the number of energy steps. According to Renyi, it is the variable scale by which the dimension is defined [41]. In Fig. 1.15 the fractal dimensions $d_f(C_x)$ are evaluated for the sputtered species from $C_1$ to $C_4$. The two parameters $I(C_x)$ and $d_f(C_x)$ are employed together to identify, distinguish and quantify the information and the fractal dimensional space generated by the externally induced dynamic processes in irradiated carbon nanotubes. The above mentioned four (4) experimental observations from the mass spectrometric data and its information-theoretic description based on $I(C_x)$ and $d_f(C_x)$ configures the two dynamic processes; the linear, cascades of the outward expanding atomic collisions of primary and secondary collision partners as CC and the nonlinear, localized hot spots of the LTS. Two different energy scales are associated with these processes that may occur in the similar spatial environments nut occur with different time scales, as will be discussed later.

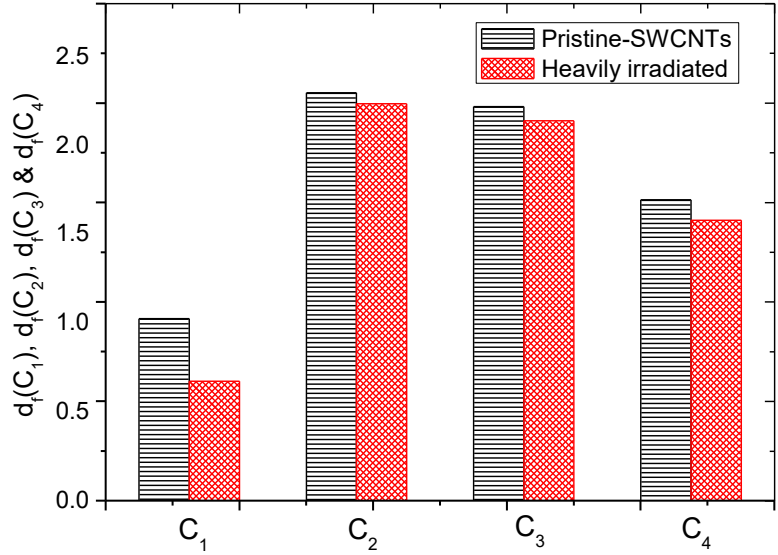

**Figure 0.15:** The fractal dimensions of the four emitted species $C_1$, $C_2$, $C_3$ and $C_4$ are plotted for the two sets of the pristine and damaged SWCNTs. From ref. [40].

The qualitative picture of $C_1$'s dependence on $Cs^+$ energy and the non-dependence of the emission of $C_2$, $C_3$ and $C_4$ on $E(Cs^+)$ is obvious from the $p(C_x)$ versus $E(Cs^+)$ plots in Fig. 1.14(a) and (b). Similar message is conveyed by the entropy plots in 1.14(c) and (d). The quantitative analysis is presented in Fig. 1.15. For $C_1$ the fractal dimension $d_f(C_1) \sim 1$ for the pristine and < 1 for the damaged SWCNTs. Fractal dimension is derived in equation (6) from the information $I_1 \equiv I(C_1)$ that emerged from the energy dissipation mechanisms with $E(Cs^+)$ dependence. The pattern of $I_1$ as a function of $E(Cs^+)$ and $d_f(C_1) \lesssim 1$ is indicative of linear, non-space-filling, collision cascades. A recent experimental investigation [46] that utilized and compared their data with the Monte Carlo simulations of SRIM [47] has confirmed the fractal dimension of monovalent vacancies $d_f(single-vacancies) \sim 1$. Our experimentally determined fractal dimension for $C_1$ is shown in figure 1(f) $d_f(C_1) \sim 1$.

For $C_2$ and higher clusters, the fractal dimension $d_f(C_{x\geq2})\sim2$ shown in Fig. 1.15. The fractal dimension $\sim2$ implies a space-filling, localized region that emits diatomic and larger clusters. Its space-filling character contrasts the linearly spreading, non-space-filling character of CCs. It can happen only if the local temperature is high $\sim T_s$ and the energies of formation of multi-vacancies are less than the energy required for a monovalent vacancy $E_{xv}[x \geq 2] < E_{1v}$. This is the essential requirement for $p(C_2) > p(C_1)$. Only a space-filling, multifractal, localized thermal spike can describe $d_f(C_2)\sim d_f(C_3)\sim2$. The conditions that generate such a spike in irradiated SWCNTs have been discussed in the preceding section. The probability of emission of a cluster $C_x$ with energy of formation $E_{xv}$ for an x-valent vacancy formation at temperature $T_s$ has the dependence of the form $p(C_x) \propto (exp(E_{xv}/T_s) + 1)^{-1}$. Here, $p(C_x)$ for a particular cluster $C_x$ is not dependent on $Cs^+$ energy but on $T_s$. The $p(C_x)\ vs\ E(Cs^+)$ spectra in Fig. 1.14(a) and (b) confirm this conclusion. For the highly irradiated and damaged SWCNTs $d_f(C_1) < 1$. This may be the result of disruption of the C-C bonding due to the creation of a significant number of 2-, 3- and 4-atom vacancies on the hexagonal network. On the other hand, the spike-related multi-atomic cluster emissions are equally proficient in damaged SWCNTs as can be seen in Fig. 1.14(b) and 1.14(d).

### 1.6.3  *Kullback-Leibler divergence or Relative entropy:*

Another important parameter, Kullback-Leibler divergence [48] or the relative entropy $D\big(p(C_x)\ \|\ p(C_y)\big)$ that can be defined by using the probability distributions of any of the two emitted species $C_x$ and $C_y$ to estimate the information-theoretic divergence between the underlying physical processes [48]. Relative probability is a measure of the divergence or distance between the probability distributions $p(C_x)$ and $p(C_y)$. In physical terms, it implies that the two sets of information generating dynamical systems that may have different origins yet may have some physical connection [48]. The CC and LTs are an example of such dynamical systems. The relative entropy is defined as

$$D(p(C_x) \| p(C_y)) = p(C_x) \ln(p(C_x)/p(C_y)) \qquad (7)$$

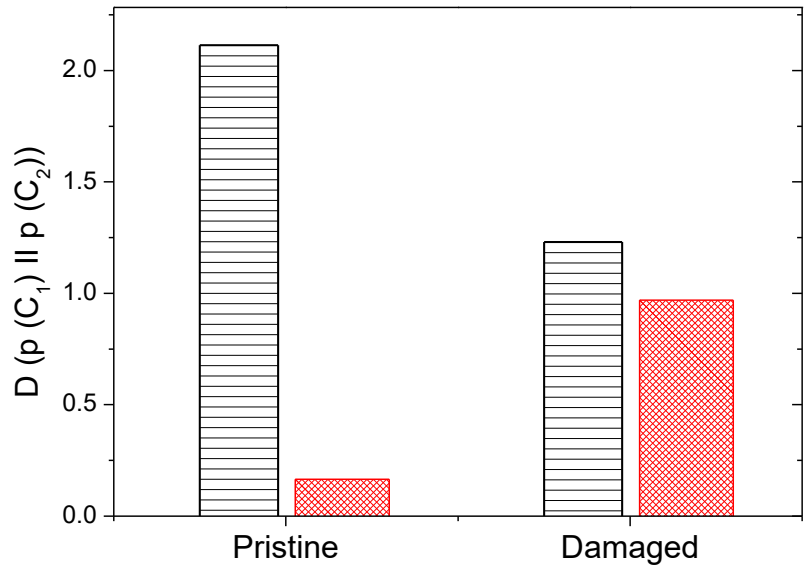

**Figure 0.16:** Relative entropy $D(p(C_1) \| p(C_2))$ as a measure of the distance between the two probability distributions of $C_1$ and $C_2$ is calculated and shown for the pristine and damaged SWCNTs [40].

In Fig. 1.16 the relative probability of $C_1$ versus $C_2$ reduces by 50% for the damaged SWCNTs as opposed to the value for the pristine nanotubes. It implies that the distance between the two probability distributions $p(C_1)$ and $p(C_2)$ and reduces in the damaged SWCNTs.

Relative probability is generally asymmetric between the two different probability distributions implying $D(p_x \| p_y) \neq D(p_y \| p_x)$. Fig. 1.16 shows that this is true for $D(p(C_1) \| p(C_2))$ and $D(p(C_2) \| p(C_1))$. This is because the origin of the emitted species

are two different physical processes that generate probability distributions which do not relate to the identical mechanisms of energy dissipation. This asymmetric relationship can be employed for identifying the underlying dissipative structures. It can be shown that whenever the origin of the two emitted species is same then the relative entropies are approximately symmetric, for example $D(p(C_2) \parallel p(C_3)) \approx D(p(C_3) \parallel p(C_2))$. It is due that $C_2^-$ and $C_3^-$ are being generated by the same thermal mechanism of LTS in the irradiated CNTs, single- or multi-walled.

It needs to be emphasized that the information theoretic entropy $I_x$ is a measure of ignorance, therefore, the relative entropy $D(p_x \parallel p_y)$ may provide relative information for the experimental outputs of the two mechanisms, which in the case of the linear cascades and nonlinear spikes are not be directly related. Their representative probability distributions have significant divergence from each other as compared with those of the set of two clusters, like $C_2$ and $C_3$ that are emitted from LTS. This, relative divergence is shown to be farther in pristine SWCNTs as opposed to that in the damaged ones.

### 1.6.4  *Spatially coherent and temporally divergent CCs and LTSs:*

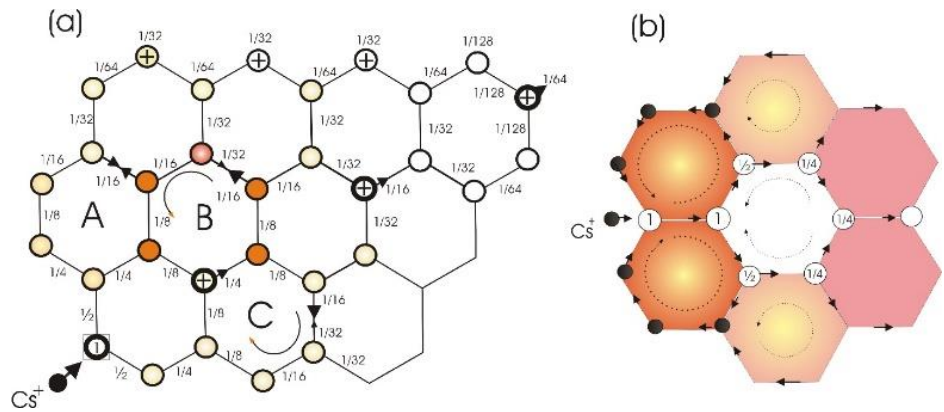

**Figure 0.17:** The origin and modular descriptions of the CCs and LTS in SWCNTs. (a) Hexagonal networks are shown with the first struck atom labelled as 1 and boxed. The cycle of the binary, atomic energy sharing collisions of neighboring atoms is shown with the received energies along the bonds. Energy recycling is shown in the rectangles A, B and C. Symbols and arrows are described below. (b) Recycling of energy induces higher vibrational temperature. Color of the hexagons indicates rising temperature [40].

Fig. 1.17(a) highlights the essential features of energy sharing by the $sp^2$-bonded atoms of the hexagons. The fundamental difference between the energy dissipation mechanisms in the typical metallic bulk solids and the $sp^2$-bonded CNTs is due to the covalently bonded carbon atoms in hexagons on a mono-shelled surface. The figure shows the hexagonal network of atoms. The first atom receives energy $E_1 \equiv 1$. If $E_1 \gg E_{dis}$—the energy that is required to break all bonds and to force the struck atom to leave its lattice site, then a binary collision cascade is generated with the energetic primary knock-on atom. The collisions proceed until the recoiling energy $<E_{dis}$. This is the well-known binary collision cascade theory. It is also the basis of Monte Carlo simulations SRIM [47]. Each collision generates a Frankel pair of interstitial atom and a vacancy. However, the concept of interstitials is valid only in bulk solids, these become the sputtered atoms in the case of SWCNTs.

For the cases where $E_1 \lesssim E_{dis}$, there may be at least three processes of energy sharing among the neighboring atoms of the extended network of hexagons. The first of these possible processes is indicated in Figure 1.17(a) by the plus (+) sign on the three atoms that are shown with arrows. The colliding atoms receive successively decreasing amounts of energy $\sim \left(\frac{1}{2}\right)^n$ with $n \geq 2$. This is the energy spreading outwards and away from the point of initial impact. The second mechanism is shown by the opposing arrows, that resembles a bow-tie, in the hexagon labelled A. In this case, the atoms on the opposing edges of the hexagon receive equal energies, in opposite directions. These collisions increase the vibrational frequencies by localizing energy in collisions with the surrounding atoms.

These collisions localize the deposited energy rather than spread it outwards. The third mechanism is shown by two unequal, opposing arrows on the interatomic bonds in hexagons B and C. These are due to collisions between atoms which deliver, or share, different amounts of energies, for example 1/16 and 1/32 in the two cases. The excess energy is directed towards the atom with lesser energy. It is further recycled in hexagons B and C as shown by curved arrows. The latter two processes increase the vibrational energies and consequently the local temperature. It eventually generates local disorder with average atomic energies $\sim \frac{1}{5}$ $to$ $\frac{1}{3} eV$. The spike temperatures $T_s \sim 3500K$ can be generated and have been calculated from the sputtered cluster probabilities [29]. At these temperature clusters are emitted from the localized subliming regions. Depending upon the initial energy, any of the atoms shown with (+) sign or arrows, can be sputtered. Whenever that happens, atomic sputtering is proposed to occurs and identified as $C_1$ peak in the mass spectrum in Figure 1(a). The experimentally observed $C_1$ yields have the direct dependence on ion energy. Figure 1.18(b) shows the recycling of the residual energy within a set of seven hexagons.

Figure 1.18 is the schematic representation of the initiation of the energy dissipative processes in the irradiated sp²-bonded atoms of the hexagonal networks of SWCNTs and MWCNTs. Different energy and time scales are associated with the two processes. The first shown in Fig. 1.17(a) is initiated with energies received by the atoms of the hexagons $E_{cascades} \geq E_{disp}$, where $E_{disp}$ is the energy required to displace an atom from its site. In SWCNTs and MWCNTs, $E_{disp} \sim 30 - 40\ eV$. Binary atomic collision cascades are initiated with typical energies $\geq E_{disp}$ and occur at time scales $\sim 10^{-15} - 10^{-14}\ s$; the lower limit depends upon the ionic energy and the higher on the primary knock-ons.

Thermal spikes are initiated with energies $\ll E_{disp}$ where the sharing and recycling of energy in the localized hexagonal patches as shown in Fig. 1.17(b). It start around $10^{-13}$ s and subsides into collective atomic vibration time scale $\sim 10^{-12}$ s. The two processes occur in the same spatial regions but happen at different time scales. Emission of $C_1$ from the collision cascades in Figure 1.17(a) represents a non-equilibrium, linear process with $d_f(C_1) \sim 1$ that occurs at RT~300K. It is the by-product of a high energy ($\geq E_{disp}$), low information-theoretic entropy, dissipative structure with durations $\sim 10^{-14}$ s.

## 1.7 Conclusions

The experimental observations from the mass spectra of the sputtered atoms and clusters from SWCNTs and MWCNTs for the different ranges of Cs$^+$ energies had shown that

(i) The domination of $C_2$, $C_3$, $C_4$ and higher clusters is a persistent feature of the mass spectra over the monatomic emissions.

(ii) The absence of $C_1$ in the mass spectra at very low Cs$^+$ energies and its consistent, low relative intensities as compared with those of the clusters, is indicative of the low efficiency of CC as opposed to the LTS.

(iii) There is an energy dependence of the probability $p(C_1)$ for the emission of $C_1$ on Cs$^+$ energy $p(C_1) \propto E(Cs^+)$, while the normalized probabilities of emission of $C_2$, $C_3$ and $C_4$ show an energy-independence.

By evaluating Shannon's information-theoretic entropies of all emitted species $C_x$ by utilizing their experimentally evaluated probability distributions $p(C_x)$ as a function of $E(Cs^+)$, their respective fractal dimensions $d_f(C_x)$ can be evaluated. The fractal dimensions of the individual species $d_f(C_x)$ can be employed to identify, distinguish and specify the externally induced dynamic processes and the associated dimensions in the irradiated CNTs. The experimental observations of the mass spectrometric data and the information-theoretic fractal dimensions $d_f(C_x)$ configure the emergence of the dynamic processes; (a) the linear CC and (b) the nonlinear LTS.

Figure 1.18 graphically demonstrates the initiation of energy sharing processes in the irradiated sp$^2$-bonded atoms in hexagonal networks of single and multi-walled carbon nanotubes and graphene sheets of nanoscale graphite powder. Two different energy scales are associated with the two processes. The first is started with energies received by the atoms of the hexagons $E_{cascades} \geq E_{disp}$, where $E_{disp}$ is the energy required to displace and atom from its site. In SWCNTs and MWCNTs $E_{disp} \sim 30 - 40\ eV$. The energy needed to generate thermal spikes $E_{spike} \ll E_{disp}$. The two processes occur in the same spatial regions but happen at different time scales. Binary atomic collision cascades are initiated with energies $\geq E_{disp}$ and occur at time scales $\sim 10^{-15} - 10^{-14}\ s$. Localized thermal spikes are initiated with energies $\ll E_{disp}$ where the sharing and recycling of energy in the hexagonal patches. It start around $10^{-13}$s and subside into the collective atomic vibration time scale $\sim 10^{-12} - 10^{-11}\ s$. Shannon entropy and the relative entropy of the probability distributions belonging to the two different physical processes and the data on fractal dimensions of the emitted species has been used here to determine the physically distinguishable features of the cascades and spikes. The graphical representation of the route to the generation of hot, subliming patches that emerge due to the energy recycling hexagons. These lead to the localized thermal spikes where $T_s \sim 3500 - 4000 K$ [29,30].

In conclusion, an information-theoretic model for the fragmenting, irradiated CNTs can describe two different yet related dissipative structures that had a common origin in the irradiating Cs ion. It can be presented with the $Cs^+$ energy as the input parameter or the agent for the dynamic events. The energy is consumed and dissipated in linear and nonlinear processes that are initiated in the mono- and multi-shelled carbon nanotubes. The output signal is in the form of sputtered atoms and clusters, emitted from the surfaces of the irradiated nanostructures. From the normalized number densities of the sputtered species the probability distributions $p(C_x)$ are constructed for every species that is emitted from the irradiated hexagonal surfaces. This information is compiled for each of the emitted constituent for the three allotropes in the form of information or Shannon entropy. Information is further employed to calculate the fractal dimension of all sputtered species. Relative entropy is calculated for pairs of the emitted species. We have shown and emphasized that fractal dimension, on its own, is a powerful analytical tool to determine the spatial nature of the physical dimension of a dissipative structure. However, under certain circumstances, it may need relative entropy to clarify the nature of the external ion-induced dynamical processes in carbon's nanostructures. Together, these information-theoretic parameters are shown to identify, distinguish and characterize the existence and operational efficiency of the linear cascades and nonlinear thermal spikes in irradiated carbon nanotubes. The model can be extended to the emergence of dissipative structures in various physical and chemical environments. This model has been employed to diagnose and ascertain the nature of the self-organizing, dynamical systems as demonstrated in the case of the emergence of Buckyball from the fragmenting fullerenes in the hot carbon soot [49, 50].

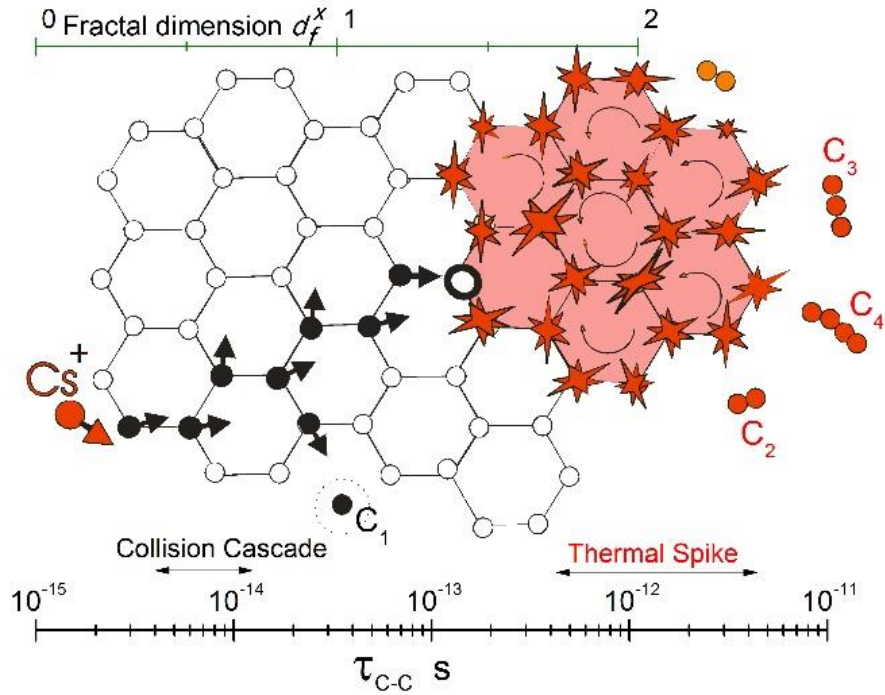

**Figure 0.18:** $Cs^+$-initiated dissipative processes in hexagonal patches. (a) For energy received by atom $>E_{disp}$ cascades are likely to start, creating single vacancies, sputtering atoms $C_1$. Probability of emission depends on linear energy dissipation $(dE/dx)^\gamma$ where $\gamma$ identifies the atomic collision mechanisms. The collision times $\tau_{C-C} \sim 10^{-14}$ s; $d_f(C_1) \sim 1$ (b) Thermal spikes generated by recycling of energy by atoms of the adjacent hexagons for $\sim 1$ eV, leading to localized spikes. Probabilities of cluster emission $p(C_x) \propto (exp(E_{xv}/T_s) + 1)^{-1}$, depend upon spike temperature $T_s$. Spikes occur for longer times $\tau_{C-C} \sim 10^{-12}$ s; fractal dimension of the representative cluster $C_2$ is $d_f(C_2) \sim 2$.

## Acknowledgements

Authors are indebted to B. Ahmad, M. N. Akhtar, A. Qayyum, T. Riffat, W. Arshad, S. A. Janjua, S. Zeeshan, S. D. Khan, A. Ashaf, K. Yaqub, S. Abbas, M. Yousuf, A. Naeem, A. Mushtaq and M, Shahnawaz for participation and support in numerous experiments, over the last two decades, reported in this chapter. The information-theoretic model employed here, was presented at PIEAS, Islamabad, where one of the authors (SA) gave a course 'Fractals and Irradiated Solids' in the spring of 2018. S. M. Mirza, M. Y. Hamza, S. Qamar, M. Ikram and M. T. Siddiqui and colleagues and students at PIEAS are acknowledged for their support and helpful discussions during the course.